# Programmable Logic Arrays

Issam W. Damaj, Dhofar University

## Introduction

Programmable Logic Arrays (*PLAs*) are widely used traditional digital electronic devices. The term "digital" is derived from the way digital systems process information; that is by representing information in digits and operating on them. Over the years, digital electronic systems have progressed from vacuum-tube circuits to complex integrated circuits, some of which contain millions of transistors. Nowadays, digital systems are included in a wide range of areas, such as, communication systems, military systems, medical systems, industrial control systems, and consumer electronics.

Electronic circuits can be separated into two groups, digital and analog circuits. Analog circuits operate on analog quantities that are continuous in value, while digital circuits operate on digital quantities that are discrete in value and limited in precision. Analog signals are continuous in time besides being continuous in value. Most measurable quantities in nature are in analog form, for example, temperature. Measuring round the hour temperature changes is continuous in value and time, where the temperature can take any value at any instance of time with no limit on precision but the capability of the measurement tool. Fixing the measurement of temperature to one reading per an interval of time and rounding the value recorded to the nearest integer will graph discrete values at discrete intervals of time that easily could be coded into digital quantities. From the given example, an analog-by-nature quantity could be converted to digital by taking discrete-valued samples at discrete intervals of time and then coding each sample. The process of conversion is usually known as analog-to-digital conversion (A/D). The opposite scenario of conversion is also valid and known as digital-to-analog conversion (D/A). The representation of information in a digital form has many advantages over analog representation in electronic systems. Digital data that is discrete in value, discrete in time, and limited in precision could be efficiently stored, processed and transmitted. Digital systems are said practically to be more noise immune as compared to analog electronic systems due to the physical nature of analog signals. Accordingly, digital systems are more reliable than their analog counter part. Examples of analog and digital systems are shown in Figure 1.

# A Bridge between Logic and Circuits

Digital electronic systems represent information in digits. The digits used in digital systems are the *0* and *1* that belong to the *binary* mathematical number system. In logic, the *0* and *1* values correspond to *True* and *False*. In circuits, the *True* and *False* could be thought of as *High* voltage and *Low* voltage. These correspondences set the relations among logic (*True* and *False*), *binary* mathematics (*0* and *1*), and circuits (*High* and *Low*).

Logic, in its basic shape, deals with reasoning that checks the validity of a certain proposition - a proposition could be either *True* or *False*. The relation among logic, *binary* mathematics, and circuits enables a smooth transition of processes expressed in propositional logic to *binary* mathematical functions and equations (*Boolean* algebra), and to digital circuits. A great scientific wealth exist that supports strongly the relations among the three different branches of science that lead to the foundation of modern digital hardware and logic design.

*Boolean* algebra uses three basic logic operations *AND*, *OR*, and *NOT*. The *NOT* operation if joined with a proposition *P* works by negating it; for instance, if *P* is *True* then *NOT P* is *False* and vice versa. The operations *AND* and *OR* should be used with two propositions, for example, *P* and *Q*. The logic operation *AND*, if applied on *P* and *Q* would mean that *P AND Q* is *True* only when both *P* and *Q* are *True*. Similarly, the logic operation *OR*, if applied on *P* and *Q* would mean that *P OR Q* is *True* when either *P* or *Q* is *True*. Truth tables of the logic operators *AND*, *OR*, and *NOT* are shown in Figure 2.a. Figure 2.b shows an alternative representation of the truth tables of *AND*, *OR*, and *NOT* in terms of *0s* and *1s*.

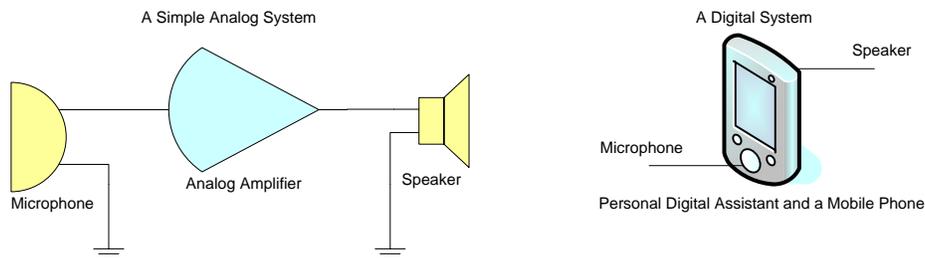

**Figure 1. A simple analog system and a digital system; the analog signal amplifies the input signal using analog electronic components. The digital system can still include analog components like a speaker and a microphone, the internal processing is digital.**

| Input *X* | Input *Y* | Output: *X AND Y* | Input *X* | Input *Y* | Output: *X OR Y* | Input *X* | Output: *NOT X* |
|---|---|---|---|---|---|---|---|
| False | False | False | False | False | False | False | True |
| False | True | False | False | True | True | True | False |
| True | False | False | True | False | True | | |
| True | True | True | True | True | True | | |

**(a)**

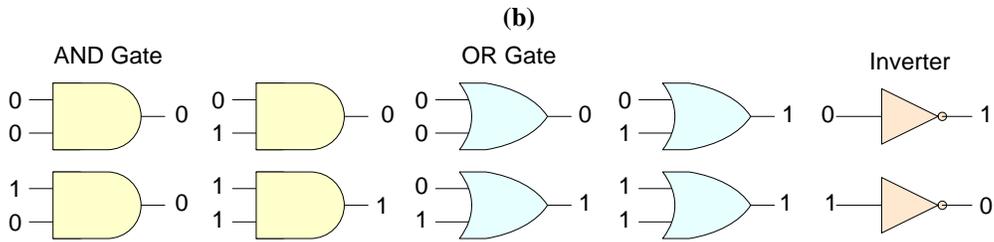

| Input X | Input Y | Output: X AND Y |
|---------|---------|-----------------|
| 0 | 0 | 0 |
| 0 | 1 | 0 |
| 1 | 0 | 0 |
| 1 | 1 | 1 |

| Input X | Input Y | Output: X OR Y |
|---------|---------|----------------|
| 0 | 0 | 0 |
| 0 | 1 | 1 |
| 1 | 0 | 1 |
| 1 | 1 | 1 |

| Input X | Output: NOT X |
|---------|---------------|
| 0 | 1 |
| 1 | 0 |

**(b)**

AND Gate  OR Gate  Inverter

**(c)**

**Figure 2. (a) Truthtables for AND, OR, and Inverter. (b) Truthtables for AND, OR, and Inverter in binary numbers, (c) Symbols for AND, OR, and Inverter with their operation.**

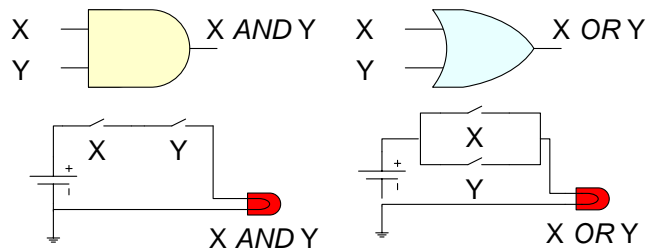

**Figure 3. A suggested analogy between *AND* and *OR* gates and electric circuits.**

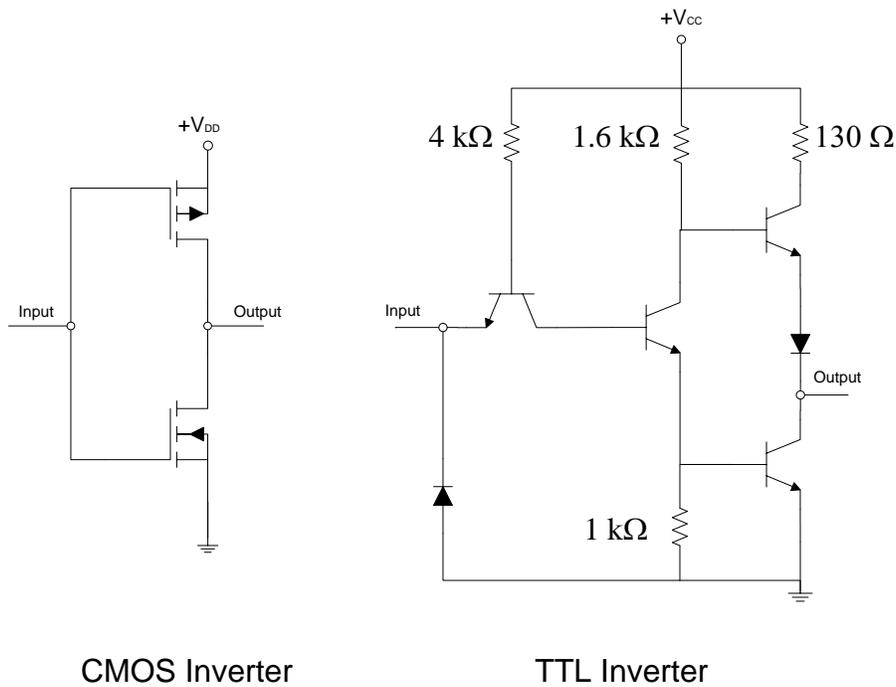

CMOS Inverter                 TTL Inverter

**Figure 4. Complementary Metal-oxide Semiconductor (*CMOS*) and Transistor-Transistor Logic (*TTL*) Inverters.**

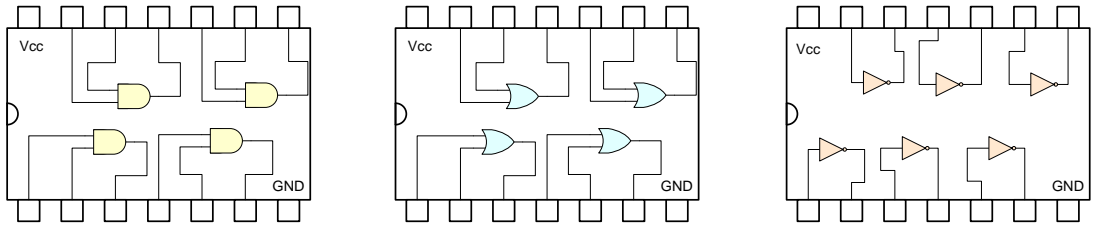

**Figure 5. The 74LS21 (*AND*), 74LS32 (*OR*), 74LS04 (*Inverter*) *TTL* ICs.**

Digital circuits implement the logic operations *AND*, *OR*, and *NOT* as hardware elements called "gates" that perform logic operations on binary inputs. The *AND*-gate performs an *AND* operation, an *OR*-gate performs an *OR* operation, and an *Inverter* performs the negation operation *NOT*. Figure 2.c shows the standard logic symbols for the three basic operations. With analogy from electric circuits, the functionality of the *AND* and *OR* gates are captured as shown in Figure 3. The actual internal circuitry of gates is built using transistors; two different circuit implementations of inverters are shown in Figure 4. Examples of *AND*, *OR*, *NOT* gates integrated circuits (ICs) are shown in Figure 5. Besides the three essential logic operations, there are four other important operations - the *NOR* (*NOT-OR*), *NAND* (*NOT-AND*), Exclusive-*OR* (*XOR*) and Exclusive-*NOR* (*XNOR*).

A logic circuit is usually created by combining gates together to implement a certain logic function. A logic function could be a combination of logic variables (such as A, B, C, etc.) with logic operations; logic variables can take only the values *0* or *1*. The created circuit could be implemented using *AND-OR-Inverter* gate-structure or using other types of gates. Figure 6 shows an example combinational implementation of the following logic function *F(A, B, C)*:

*F(A, B, C) = ABC + A'BC + AB'C'*

*F(A, B, C)* in this case could be described as a sum-of-products (*SOP*) function according to the analogy that exists between *OR* and addition (+), and between *AND* and product (.); the *NOT* operation is indicated by an Apostrophe " ' " following the variable name.

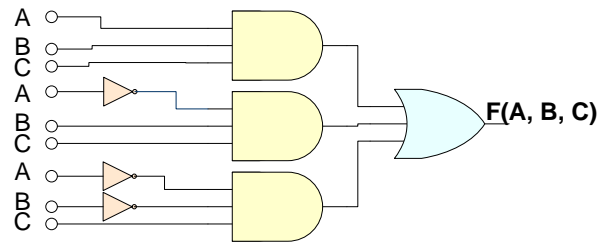

**Figure 6. AND-OR-Inverter implementation of the function *F(A, B, C) = ABC + A'BC + AB'C'*.**

## Programmable Logic

Basically, there are three types of *IC* technologies that can be used to implement logic functions on (1), these are, full-custom, semi-custom, and programmable logic devices (*PLDs*). In full-custom implementations, the designer cares about the realization of the desired logic function to the deepest details including the transistor-level optimizations to produce a high-performance implementation. In semi-custom implementations, the designer uses some ready logic-circuit blocks and completes the wiring to achieve an acceptable performance implementation in a shorter time than full-custom procedures. In *PLDs*, the logic blocks and the wiring are ready. In implementing a function on a *PLD*, the designer will only decide of which wires and blocks to use; this step is usually referred to as programming the device.

Obviously, the development time using a *PLD* is shorter than the other full-custom and semi-custom implementation options. The performance of a *PLD* varies according to its type and complexity; however, a full-custom circuit is optimized for achieving higher performance. The key advantage of modern programmable devices is their reconfiguration without rewiring or replacing components (re-programmability). Programming a modern *PLD* is as easy as writing a software program in a high-level programming language.

The first programmable device which achieved a widespread use was the Programmable Read-Only Memory (*PROM*) and its derivatives Mask-*PROM*, and Field-*PROM* (the Erasable or Electrically Erasable versions). Another step forward took place in this field which led to the development of *PLDs*. Programmable Array Logic (*PAL*), Programmable Logic Array (*PLA*), and Generic Array Logic (*GALs*) are commonly used *PLDs* designed for small logic circuits and referred to as Simple-*PLDs* (*SPLDs*). Other types, the Mask-Programmable Gate Arrays (*MPGAs*), were developed to handle larger logic circuits. Complex-*PLDs* (*CPLDs*) and Field Programmable Gate Arrays (*FPGAs*) are more complicated devices that are fully programmable and instantaneously customizable. Moreover, *FPGAs* and *CPLDs* have the ability to implement very complex computations with millions of gates devices currently in production. A classification of *PLDs* is shown in Figure 7.

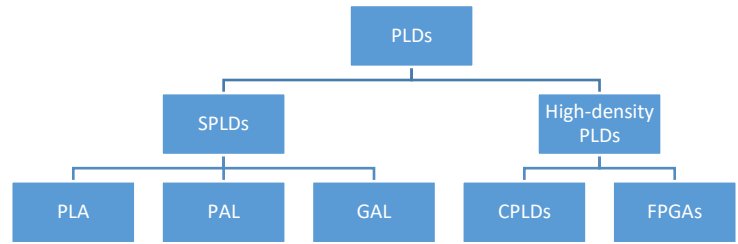

**Figure 7. Typical *PLD* devices classification.**

## Programmable Logic Arrays

A Programmable Logic Array (*PLA*) is a simple programmable device (*SPLD*) used to implement combinational logic circuits. A *PLA* has a set of programmable *AND* gates, which link to a set of programmable *OR* gates to produce an output (see Figure 8). Implementing a certain function using a *PLA* requires the determination of which connections among wires to keep. The unwanted routes could be eliminated by burning the switching device (possibly a fuse or an antifuse) connecting different routs. The *AND-OR* layout of a *PLA* allows for implementing logic functions that are in an *SOP* form.

Technologies usually used to implement programmability in *PLAs* include fuses or antifuses. A fuse is a low resistive element that could be blown (programmed) to result in an open circuit or high impedance. An antifuse is a high resistive element (initially high impedance) and is programmed to be low impedance.

Boolean expressions can be represented in either of two standard forms, the sum-of-products (*SOPs*) and the product-of-sums (*POSs*). For example, the equations for *F* (an *SOP*) and *G* (a *POS*) are as follows:

*F(A, B, C) = ABC + A'BC + AB'C'*
*G(A, B, C) = (A + B + C) . ( A' + B + C) . (A + B' + C')*

A product term is a term consisting of the *AND* (*Boolean* multiplication) of literals (*A*, *B*, *C*, etc.). When two or more product terms are summed using an *OR* (*Boolean* addition), the resulting expression is an *SOP*. A standard *SOP* expression *F(A, B, C, ...)* is one where all the variables appear in each product term. Standardizing expressions makes evaluation, simplification, and implementation much easier and systematic.

The implementation of any *SOP* expression using *AND-*gates*, OR-*gates*,* and *inverters*, could be easily replaced using the structure offered by a *PLA*. The algebraic rules of hardware development using standard *SOP* forms are the theoretical basis for designs targeting *PLAs*. The design procedure simply starts by writing the desired function in an *SOP* form, and then the implementation works by choosing which fuses to burn in a fused-*PLA*.

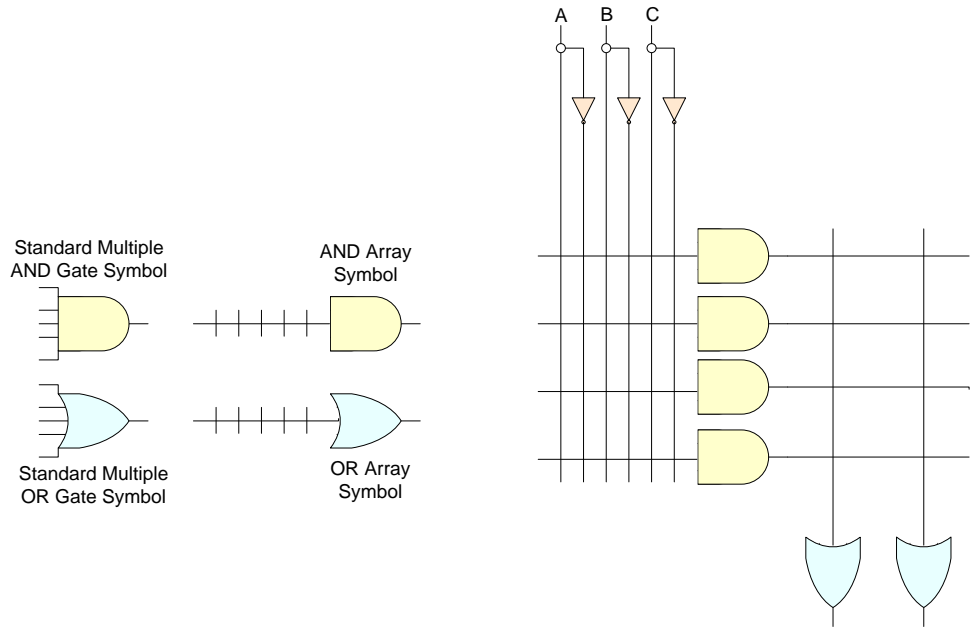

**Figure 8. A 3-input 2-output PLA with its AND Arrays and OR Arrays. An AND array is equivalent to a standard multiple-input AND gate, and an OR array is equivalent to a standard multiple-input OR gate.**

In the following two examples, we demonstrate the design and implementation of logic functions using *PLA* structures. In the first example, we consider the design and implementation of a three variables majority function. The function *F(A, B, C)* will return a *1* (*High* or *True*) whenever the number of *1s* in the inputs is greater than or equal to the number of *0s*. The truthtable of *F* is shown in Figure 9. The terms that make the function *F* return a *1* are the terms *F(0, 1, 1)*, *F(1, 0, 1)*, *F(1, 1, 0)*, or *F(1, 1, 1)*. This could be alternatively formulated as in the following equation:

$$F = A\text{'}BC + AB\text{'}C + ABC\text{'} + ABC$$

In Figure 10, the implementations using a standard *AND-OR-Inverter* gate-structure and a *PLA* are shown.

Another function G(A, B, C, D) could have the following equation:

$$G = A\text{'}B + AB\text{'}CD + BC\text{'} + ABD + B\text{'}C\text{'}D\text{'}$$

The implementation of *G* using a *PLA* is shown in Figure 11.

| Input A | Input B | Input C | Output F |
|---------|---------|---------|----------|
| 0 | 0 | 0 | 0 |
| 0 | 0 | 1 | 0 |
| 0 | 1 | 0 | 0 |
| 0 | 1 | 1 | 1 |
| 1 | 0 | 0 | 0 |
| 1 | 0 | 1 | 1 |
| 1 | 1 | 0 | 1 |
| 1 | 1 | 1 | 1 |

**Figure 9. Truthtable of the majority function.**

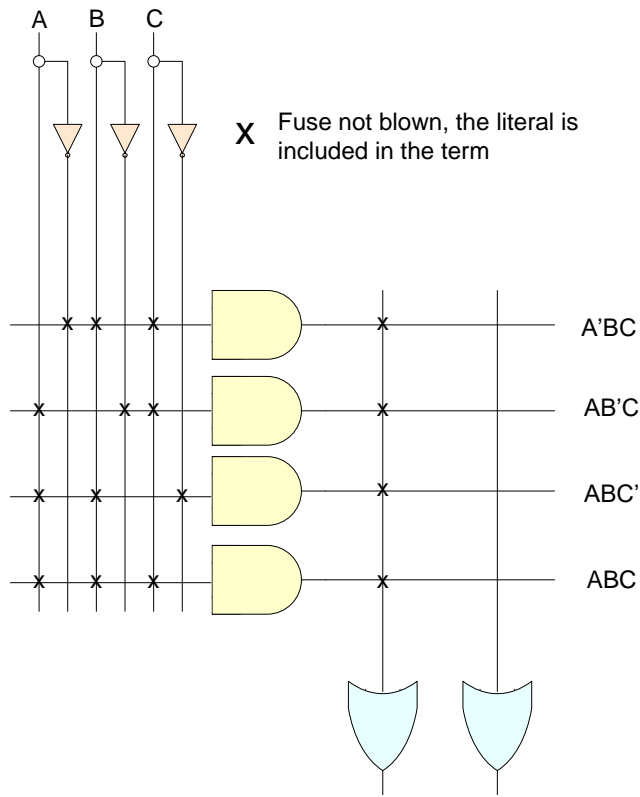

**Figure 10. *PLA* implementation of F(A, B, C).**

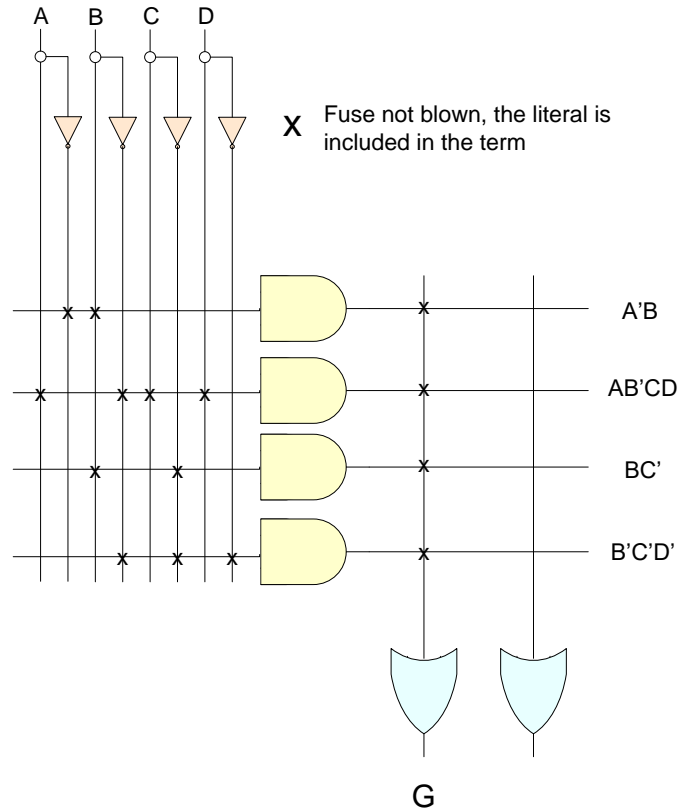

**Figure 11.** *PLA* implementation of G(A, B, C, D).

## Early PLAs

Near the beginning of 1970s, companies like *Philips*, *Texas Instruments*, *National Semiconductor*, *Intersil*, *IBM* (2) and *Signetics* introduced early *PLA* and *PLA*-based devices. Early *PLAs* had limited numbers of input/output ports (around twenty), array cells count (from hundreds to few thousands), and speeds (with around 1 to 35 nanoseconds delay). Later *PLAs* performed with greater speeds (with around 2 to 5 nanoseconds delay), with array sizes of thousands of cells, and input/output ports number of around 100 (3). Currently, some *PLA*-structures are parts of high-density, high-performance, and complex programmable logic devices (*CPLDs*).

Currently, *PLAs* are available in the market in different types. *PLAs* could be standalone chips, or parts of bigger processing systems. Standalone *PLAs* are available as Mask Programmable (*MPLAs*) and Field Programmable (*FPLAs*) devices. Mask Programmable *PLAs* are programmed at the time of manufacture, while Field Programmable *PLAs* can be programmed by the user with a computer-aided design (*CAD*) tool.

## PLAs in Modern Complex Systems and Areas of Application

*PLAs* have largely motivated the development of many modern programmable systems. *PLAs* are usually used as parts of more complicated processing systems. *PLAs* have also inspired the creation of complex *PLA*-based systems with *PLA*-like structure. The available variety of *PLAs* and *PLA*-based systems paved the way for their employment in many areas of application.

The *CoolRunner II CPLD* from *Xilinx* uses a *PLA* type structure. This device has multiple function blocks (*FBs*). Each function block contains sixteen macrocells, the function blocks are interconnected by an advanced interconnect matrix (*AIM*). A basic architectural block diagram for the *CoolRunner II* with a greatly simplified diagram of a function block (*FB*) is shown in Figure 12.

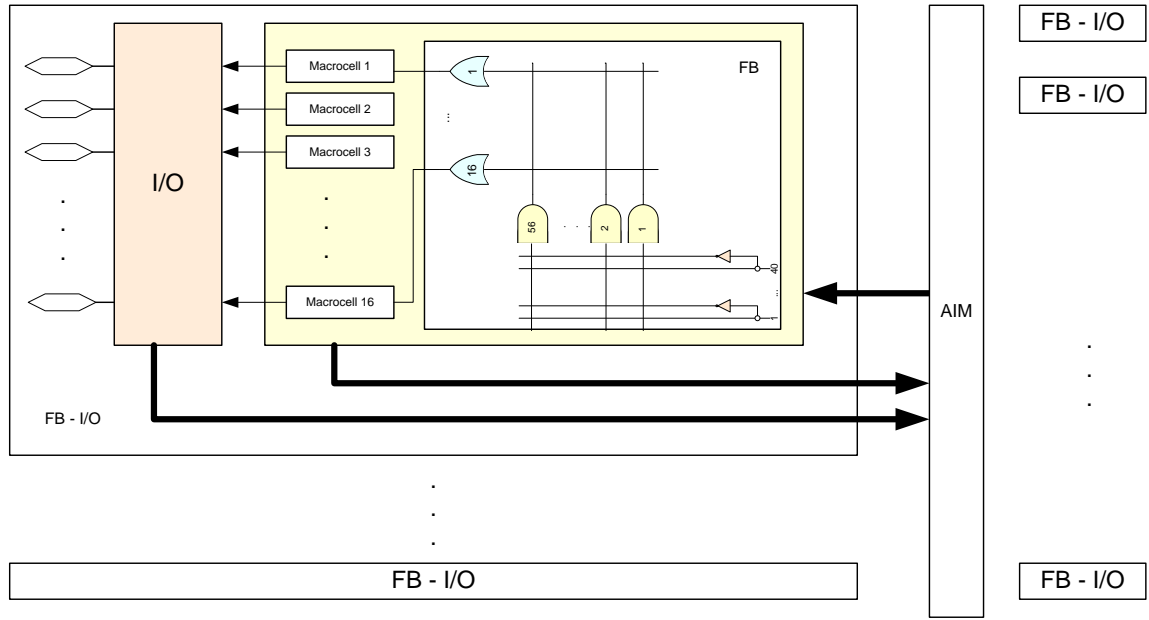

**Figure 12. Architectural block diagram for the *CoolRunner II*.**

The *CoolRunner II* series of *CPLDs* contains from 32 macrocells to 512 macrocells. The number of function blocks range from 2 to 32. The *PLA* structure contains a programmable *AND* array with *56 AND*-gates, and a programmable *OR* array with *16 OR*-gates. With the *PLA* structure, any product term can be connected to any *OR* gate to create an *SOP* output. Each *FB* can produce up to *16 SOP* outputs each with *56* product terms.

The main additions to the traditional *PLA* structure in a device like *CoolRunner II* are the complex macrocells. A macrocell can be configured for combinational logic or sequential logic (with availability of flip-flops). The macrocell in a *CoolRunner II* also contain an

*XOR*-gate to enable complementing the *SOP* output (coming from the *PLA OR*-gate) to produce a *POS* form. A *1* on the input of the *XOR*-gate complements the *OR* output (a *POS* form is produced) and a *0* keeps the output uncomplemented (in an *SOP* form). Choices between *SOP* forms and *POS* forms, various clock inputs, flip-flop include or bypass, are done using different multiplexers.

Another famous device with a *PLA*-like structure is the *ICT* Programmable Electrically Erasable Logic *(PEEL) Array* (4). *PEEL Arrays* are large *PLAs* that include macrocells with flip-flops. The *PEEL Array* structure is shown in Figure 13 with its *PLA*-like planes; the outputs of the *OR*-plane are divided into groups of four, and each group can be input to any of the logic cells. The logic cells, depicted in Figure 14, provide registers for the sum terms and can feed-back the sum terms to the *AND*-plane. The logic cells can also connect the sum terms to the Input/Output pins. The multiplexers each produce an output of the logic cell and can provide either a registered or combinational output. Because of their *PLA*-like planes, the *PEEL Arrays* are well-suited to applications that require *SOP* terms.

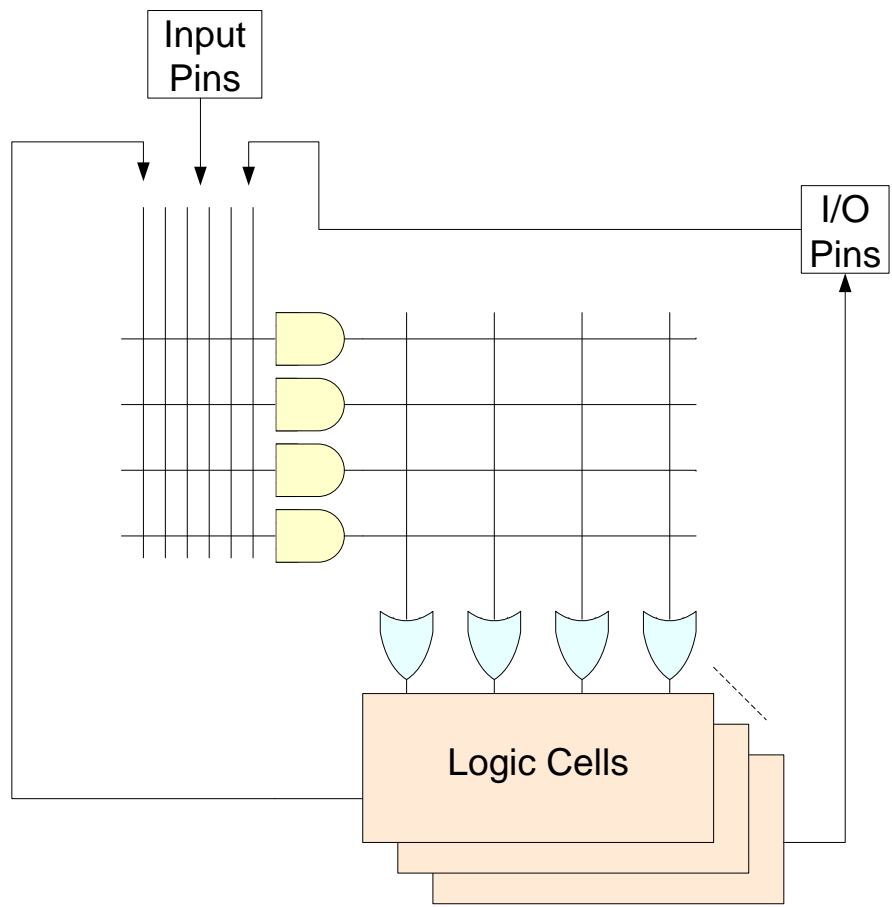

**Figure 13. Main components in the architecture of *ICT PEEL Arrays*.**

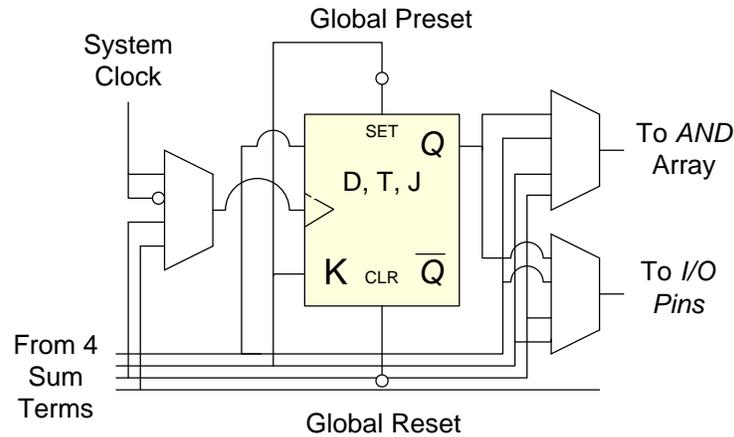

**Figure 14. Structure of *PEEL Array* Logic Cell.**

The Multiple *ALU* Architecture with Reconfigurable Interconnect Experiment System (*MATRIX*) is another modern architecture that benefits from the *PLA* architecture (5). The *MATRIX* architecture is unique in that it aims to unify resources for instruction storage and computation. The basic unit (*BFU*) can serve either as a memory or a computation unit. The *8 BFUs* are organized in an array, and each *BFU* has a *256*-word memory, *ALU*-multiply unit and reduction control logic. The interconnection network has a hierarchy of three levels; it can deliver up to *10 GOPS* with *100 BFUs* when operating at *100* MHz. The *MATRIX* controller is composed of a pattern matcher for generating local control from the *ALU* output, a reduction network for generating local control, and a *20*-input, *8*-output *NOR* block which serve as half of a *PLA*.

One famous application of *PLAs* is to implement the control over a datapath in a processor. A datapath controller usually follows predefined sequences of states. In each control state, the *PLA* part of the controller will determine what are the datapath control signals to produce and what is the next state of the controller. The design of the controller usually starts by formulating different states and transitions using a state diagram. The state diagram is then formulated in a truthtable form (state transition table), where *SOP* equations could be produced. The derived *SOP* equations are then mapped onto the *PLA*. A design example of a datapath controller is shown in Figures 15, 6, 17 and 18. Figure 16 shows a typical controller state diagram. Figure 17 depicts the block diagram of the datapath controller. Figure 18 suggests a *PLA* implementation of the controller.

Many areas of application have benefited from *PLAs* and *PLA*-based devices, such as, cryptography (6), signal processing (7), computer graphics (8), image processing (9), data mining (9), networking (10), etc.

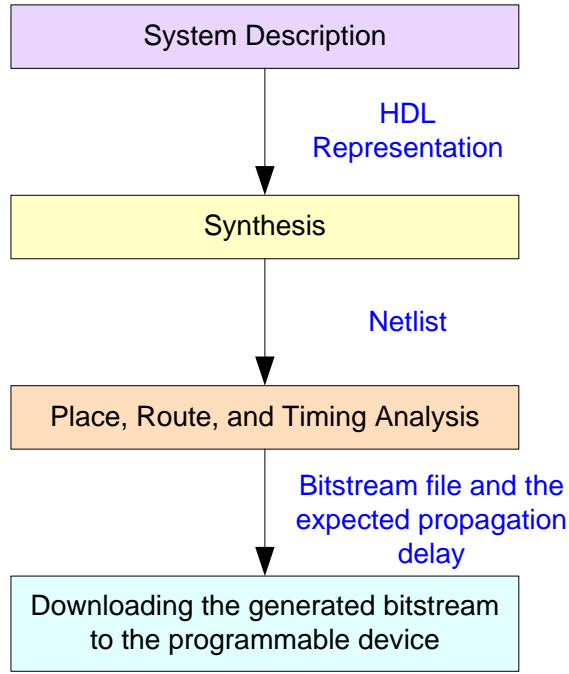

**Figure 15. The process of hardware describe-and-synthesize development for programmable logic devices.**

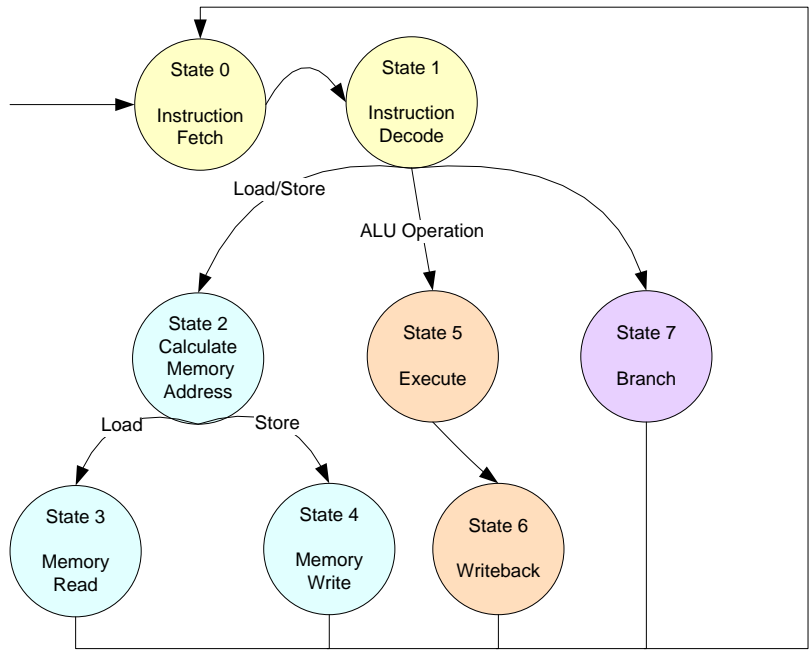

**Figure 16. A design example of a datapath controller; the State Diagram.**

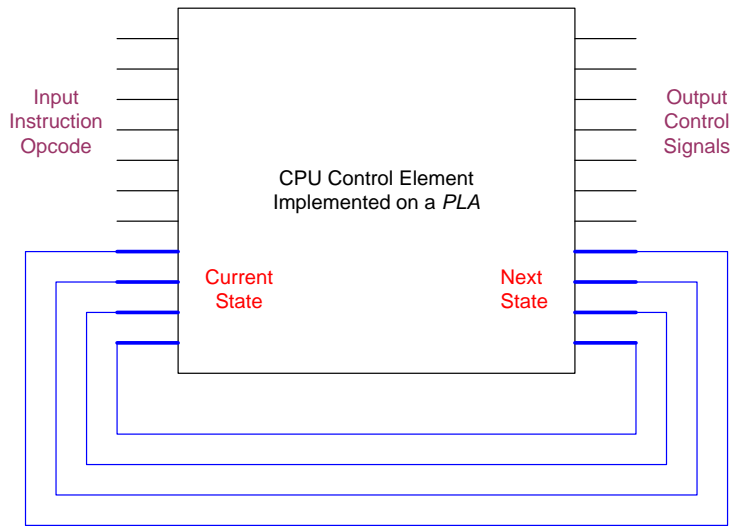

**Figure 17. A design example of a datapath controller; Control Element block diagram**

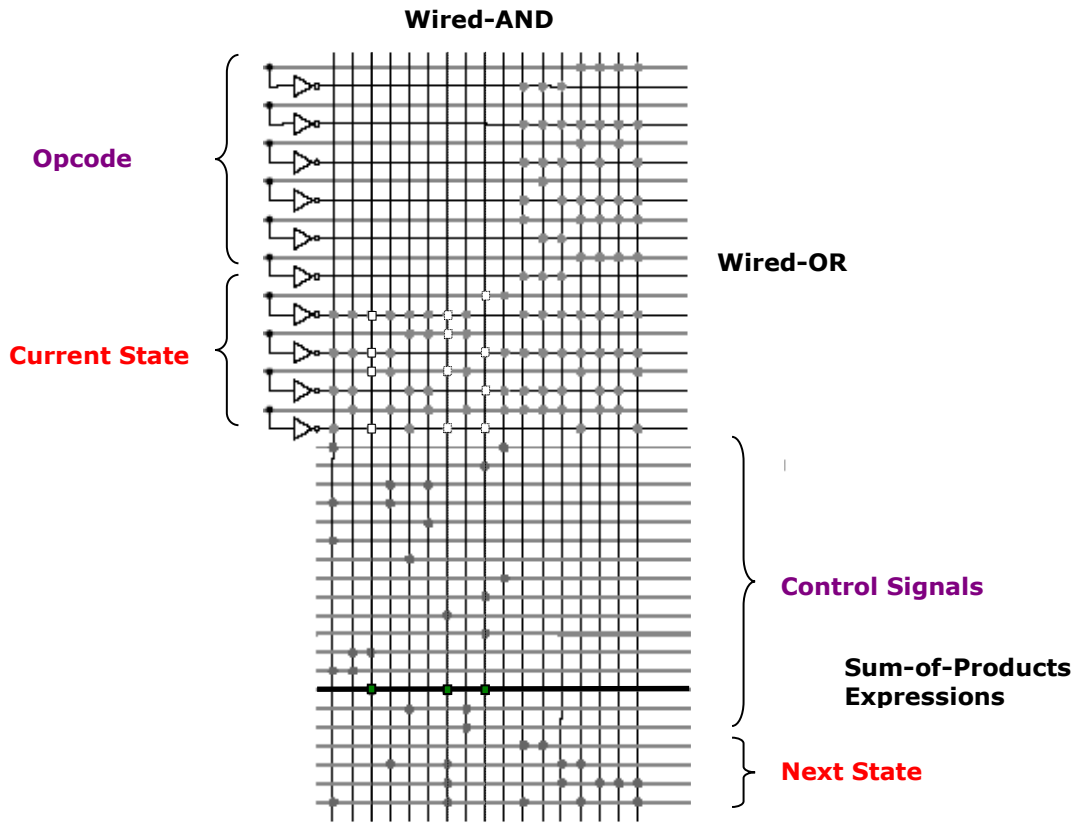

**Figure 18. A design example of a datapath controller; *PLA* internal implementation.**

## Programming PLAs

Traditional *PLAs* are usually programmed using a *PLA* device programmer (like traditional *PROMs* and *EPROM*-based logic devices). Some more complex *PLA*-based devices, such as *CPLDs*, can be also programmed using device programmers; nevertheless modern *CPLDs* are in-circuit programmable. In other words, the circuit required to perform device programming is provided within the *CPLD* chip. In-circuit programmability makes it possible to erase and reprogram the device without an external device programmer.

Modern *CPLDs*, including the internal *PLA*-like structures, benefit from the latest advances in the area of hardware/software co-design. Descriptions of the desired hardware structure and behavior are written in a high-level context using hardware description languages (*HDLs*), such as *VHDL* or *Verilog*. The description code is then compiled and downloaded in the programmable device prior to execution. Schematic captures are also an option for design entry. Schematic captures has become less popular especially with complex designs. The process of hardware describe-and-synthesize development for programmable logic devices is shown in Figure 15.

Hardware compilation consists of several steps. Hardware synthesis is the first major step of compilation, where an intermediate representation of the hardware design (called a netlist) is produced. A netlist is usually stored in a standard format called the Electronic Design Interchange Format (*EDIF*) and it is independent of the targeted device details. The second step of compilation is called place and route, where the logical structures described in the netlist are mapped onto the actual macrocells, interconnections, and input and output pins of the targeted device. The result of the place and route process is a usually called a bitstream. The bitstream is the binary data that must be loaded into the *PLD* to program it to implement a hardware design.

## The renewable usefulness of PLAs

*PLAs* and their design basis have witnessed a renewable importance and have been the choice of designers for many systems as well as the target of different design methodologies. The renewable usefulness of *PLAs* is clear from the number of investigations carried out relying based on *PLAs*.

A sub-threshold circuit design approach based on asynchronous micro-pipelining of a levelized network of *PLAs* is investigated in (11). The main purpose of the presented approach is to reduce the speed gap between sub-threshold and traditional designs. Energy saving is noted when using the proposed approach in a factor of four as compared to a traditional designed network of PLAs.

In (12) the authors propose a maximum crosstalk minimization algorithm taking logic synthesis into consideration for *PLA* structures. To minimize the crosstalk, technique of permuting wires is used. The *PLA* product terms lines are partitioned into long set and short set, and then product lines in long set and short set are interleaved. The interleaved wires are then checked for the maximum coupling capacitance for the aim of reducing the maximum crosstalk effect of the *PLA*.

A logic synthesis method for an *AND-XOR-OR* type sense-amplifying *PLA* is proposed in (13). Latch sense-amplifiers and a charge sharing scheme are used to achieve low-power dissipation in the suggested *PLA*.

Testable design for detecting stuck-at and bridging faults in *PLAs* is suggested in (14). The testable design is based on Double Fixed-Polarity Reed-Muller Expressions (*DFPRMs*). An *XOR* part is proposed in the design implemented in a tree structure to reduce circuit delay.

A *VLSI* approach addressing cross-talk problem in Deep Sub-Micron (*DSM*) *IC* design is investigated in (15). Logic netlists are implemented in the form of a network of medium sized *PLAs*. Two regular layout "fabrics" are used in this methodology, one for areas where *PLA* logic is implemented, and another for routing regions between logic blocks.

In (16), a *PLA*-based performance optimization design procedure for standard-cells is proposed. The optimization is done by implementing circuits' critical paths using *PLAs*. *PLAs* are proven to be good for such a replacement approach as they exhibit a gradual increase in delay as additional items are added. The final optimized hybrid design is to contain standard cells and a *PLA*.

A performance-driven mapping algorithm for *CPLDs* with many *PLA*-style logic cells is proposed in (17). The primary goal of the mapping algorithm is to minimize the depth of the mapped circuit. The algorithm included applying several heuristic techniques for area reduction, threshold control of *PLA* fan-outs and product terms, slack-time relaxation, and *PLA*-packing.

The attractions of *PLAs* that brought them to mainstream engineers include their simplicity, relatively small circuit area, predictable propagation delay, and ease of development. The powerful-but-simple property brought *PLAs* to rapid prototyping, synthesis, design optimization techniques, embedded systems, traditional computer systems, hybrid high-performance computing systems, etc. Indeed, there has been renewable interests in working with the simple *AND*-to-*OR PLAs*.

**Bibliography.**

## Reading List

## Cross-references

Programmable Logic Devices, See *PLDs*.
Synthesis, See High-level Synthesis
Design, See Logic Design